\documentclass[sigconf]{acmart}

\usepackage{appendix}
\usepackage{enumitem}
\newgeometry{top=1in,bottom=1in,right=0.7in,left=0.7in}

\usepackage[linesnumbered,ruled,vlined]{algorithm2e}
\usepackage{subfigure}
\usepackage{multirow}
\usepackage{balance}
\usepackage[normalem]{ulem}

\usepackage{amsfonts,amssymb}
\usepackage{amsmath,bm}
\usepackage{threeparttable}
\newtheorem{myDef}{Definition}
\usepackage{color}
\usepackage{soul}
\usepackage{float}

\SetKwInput{KwInput}{Input}                
\SetKwInput{KwOutput}{Output}              

\newcommand{\Method}[0]{\textsf{DualMatch}}

\newcommand{\Revision}[1]{#1}

\newcommand{\smallsection}[1]{\noindent \textbf{#1.}}

\newcommand{\Encoder}[0]{Dual-Encoder}
\newcommand{\Decoder}[0]{GM-Decoder}
\newcommand{\TimeEncoder}[0]{Temporal-Encoder}
\newcommand{\RelEncoder}[0]{Relational-Encoder}

\newcommand{\DICEWS}[1]{DICEWS{#1}}

\newcommand{\WIKIYAGO}[1]{WY50K{#1}}
\newcommand{\WIKIYAGOHybrid}[0]{WY20K}

\newcommand{\HitNFull}[0]{Hits@$N$}
\newcommand{\HitN}[0]{H@$N$}
\newcommand{\HitOne}[0]{H@$1$}
\newcommand{\HitTen}[0]{H@$10$}
\newcommand{\MRR}[0]{MRR}

\newcommand{\CFLine}[1]{ (cf. Line #1 of Algorithm~\ref{alg:process})}
\newcommand{\CFLines}[2]{ (cf. Lines #1-#2 of Algorithm~\ref{alg:process})}

\AtBeginDocument{%
  \providecommand\BibTeX{{%
    \normalfont B\kern-0.5em{\scshape i\kern-0.25em b}\kern-0.8em\TeX}}}

\setcopyright{acmcopyright}
\copyrightyear{2023}
\acmYear{2023}
\setcopyright{acmlicensed}\acmConference[WWW '23]{Proceedings of the ACM
Web Conference 2023}{May 1--5, 2023}{Austin, TX, USA}
\acmBooktitle{Proceedings of the ACM Web Conference 2023 (WWW '23), May
1--5, 2023, Austin, TX, USA}
\acmPrice{15.00}
\acmDOI{10.1145/3543507.3583381}
\acmISBN{978-1-4503-9416-1/23/04}

\begin{document}

\title{Unsupervised Entity Alignment for Temporal Knowledge Graphs}


\author{Xiaoze Liu}
\affiliation{%
  \institution{Zhejiang University}
  \city{Hangzhou}
  \country{China}}
\email{xiaoze@zju.edu.cn}
\author{Junyang Wu}
\affiliation{%
  \institution{Zhejiang University}
  \city{Hangzhou}
  \country{China}}
\email{wujunyang@zju.edu.cn}


\author{Tianyi Li}
\affiliation{%
  \institution{Aalborg University}
  \city{Aalborg}
  \country{Denmark}}
\email{tianyi@cs.aau.dk}

\author{Lu Chen}
\affiliation{%
  \institution{Zhejiang University}
  \city{Hangzhou}
  \country{China}}
\email{luchen@zju.edu.cn}
\author{Yunjun Gao}
\affiliation{%
  \institution{Zhejiang University}
  \city{Hangzhou}
  \country{China}}
\email{gaoyj@zju.edu.cn}


\renewcommand{\shortauthors}{Xiaoze Liu, Junyang Wu, Tianyi Li, Lu Chen, and Yunjun Gao}

\begin{abstract}
Entity alignment (EA) is a fundamental data integration task that identifies equivalent entities between different knowledge graphs (KGs).
Temporal Knowledge graphs (TKGs) extend traditional knowledge graphs by introducing timestamps, which have received increasing attention.
State-of-the-art time-aware EA studies have suggested that the temporal information of TKGs facilitates the performance of EA.
However, existing studies have not thoroughly exploited the advantages of temporal information in TKGs. Also, they perform EA by pre-aligning entity pairs, which can be labor-intensive and thus inefficient.
In this paper, we present \Method{} that effectively fuses the relational and temporal information for EA. \Method{} transfers EA on TKGs into a weighted graph matching problem. More specifically, \Method{} is equipped with an unsupervised method, which achieves EA without necessitating the seed alignment. 
\Method{} has two steps: (i) encoding temporal and relational information into embeddings separately using a novel label-free encoder, Dual-Encoder; and (ii) fusing both information and transforming it into alignment using a novel graph-matching-based decoder, GM-Decoder.
\Method{} is able to perform EA on TKGs with or without supervision, due to its capability of effectively capturing temporal information.
Extensive experiments on three real-world TKG datasets offer the insight that \Method{} 
significantly outperforms the state-of-the-art methods.

\end{abstract}



\maketitle

\section{Introduction}
\label{sec:intro}

Knowledge graphs (KGs) represent structured knowledge related to real-world objects, which plays a crucial role in many real-world applications, such as semantic search~\cite{XiongPC17}, entity~\cite{ner1, ner2, typing,typing2} and relation extraction~\cite{hu2021semi,liu2022hierarchical,hu2021gradient,hu2020selfore}.
However, existing KGs are generally highly incomplete~\cite{OpenEA2020VLDB}. Since different KGs are constructed from various data sources, they contain unique information but have overlapped entities. 
This scenario provides us an opportunity to integrate different KGs with the overlapped entities. 

\begin{figure}[t]
\centering 
\includegraphics[width=3.1in]{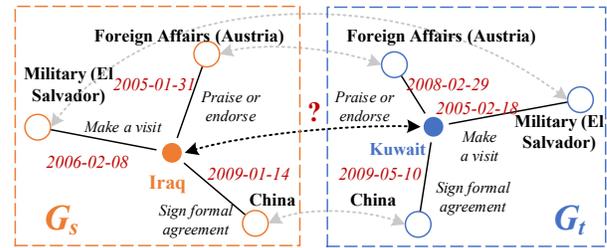} 
\vspace{-2mm}
\caption{An example of limitations of existing methods without considering temporal information}
\label{fig:example_tkg}
\vspace{-6mm}
\end{figure}

A typical strategy for integration of KGs is \textbf{Entity Alignment} (EA)~\cite{OpenEA2020VLDB}, which aligns entities from different KGs that refer to the same real-world objects. Given two KGs and a small set of pre-aligned entities (also known as \emph{seed alignment}), EA identifies all possible alignments between them. 
Existing embedding-based studies~\cite{KECG19, RREA20, AliNet20, HyperKA20, DualAMN21} have proven highly effective to perform EA, which is mainly benefited by the use of Graph Neural Networks (GNNs)~\cite{GCN17, GAT18, GraphSAGE17}. 
They assume that the neighbors of two equivalent entities in separated KGs are also equivalent~\cite{AttrGNN20}. Based on this assumption, they align entities by applying representation learning to KGs. We summarize the process of Embedding-based EA as the following two steps: 
(i) encoding entities of two KGs into embedding vectors by training an EA model with some \emph{seed alignments}; and 
(ii) decoding the embedding vectors of the entities into an \emph{alignment matrix}, based on a specific similarity measurement (e.g., cosine similarity) of their embeddings.

Most of the existing EA studies need to pre-align training sets before an embedding model can be trained, which is labor-intensive and degrades their usability in real-world applications~\cite{OpenEA2020VLDB}. These problems can be tackled by name-based~\cite{AttrGNN20, LargeEA22, CEAFF20, SEU21, EASY21,wang2022promptem, ge2021collaborem} or image-based~\cite{EVA20} models via incorporating side information, 
where few or no label is required for performing EA.
However, such models still suffer from problems. First, the performance of the name-based models is overestimated due to name-bias~\cite{JEANS20, AttrGNN20, EVA20, NoMatch21}. In this case, \Revision{the ground truths (inter graph/language links) are generally generated with the entity name, and thus results in test data leakage when they are served as features.}
Second,
the images are not inherent parts of KGs and are thus hard to be collected. 
As a result, incorporating side information may not be able to enhance EA. Moreover, most of the existing EA methods~\cite{MTransE17,MuGNN19, JEANS20,  MRAEA20, RREA20, SEU21,  DualAMN21, KECG19, EVA20, AttrGNN20, DATTI22} do not consider temporal information. This results in incorrect predictions on entities, which have similar neighborhood structures in relational triples but correspond to different time intervals, as shown in Figure~\ref{fig:example_tkg}.


%

Temporal information are often utilized to enhance the performance of various applications~\cite{li2020compression, li2021trace}.
Temporal Knowledge Graphs (TKGs)~\cite{YAGO, Wikidata14}, a type of KGs, have been proposed that associates relational facts with temporal information. As the formats of storing temporal information in TKGs are almost identical, the alignment information can be obtained easily, which naturally enhances the EA performance. Nevertheless, recent studies~\cite{TEAGNN21, TREA22} that applies TKGs to EA 
 still suffer from the following two problems.

\begin{itemize}[topsep=0pt,itemsep=0pt,parsep=0pt,partopsep=0pt,leftmargin=*]
\item \textbf{``Redundantly'' pre-align\Revision{ed} seeds.}
   They~\cite{TEAGNN21, TREA22} follow studies that applies KGs to EA, which cannot perform EA until the pre-aligned seeds are obtained. However, unlike KGs,
    temporal information in relational triples of TKGs is naturally aligned since they represent real-world time points or time periods~\cite{TEAGNN21, TREA22}.
    This character of TKGs provides the opportunity of developing time-aware EA methods in an unsupervised fashion by treating temporal information as seed alignment. 
    \item  \textbf{Insufficient temporal information discovery.} 
    They~\cite{TEAGNN21, TREA22} simply create
    embeddings for each timestamp, and treat the time intervals as relation types to enhance the graph learning process. These processes highly rely on the seed alignment, where the temporal information has not been fully exploited, and thus the accuracy of EA is limited. We use an example to illustrate it.
\end{itemize}

\begin{figure}[t]
\centering 
\includegraphics[width=3.5in]{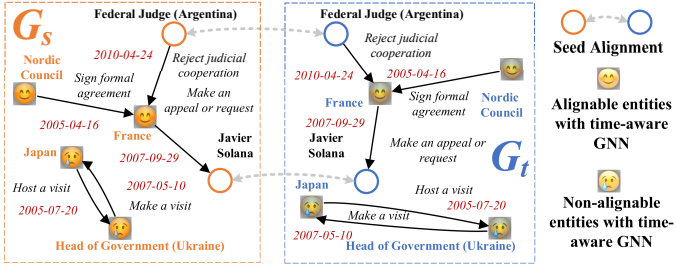} 
\vspace{-4mm}
\caption{A corner case of possible misalignment for previous time-aware GNN studies }
\vspace{-6mm}
\label{fig:example}
\end{figure}

\vspace{-2mm}
\begin{example}
\label{exp:corner_case} 
Assume that we have trained an accurate time-aware GNN on these graphs that could maximally propagate all the alignment information, as depicted in Figure~\ref{fig:example}. 
In this case, the entity ``Japan'' is likely to be misaligned due to two reasons: (1) ``Japan'' and the seeds are in different connected components, where the GNNs is unlikely  to propagate the alignment information to the embedding of ``Japan''; and (2) ``Japan'' does not share common timestamps and relation types with the seeds, and thus, the information of seeds is unable to reach ``Japan'' via shared temporal or relational embeddings.
This misalignment of Japan renders it is excluded from the objective of the learned model. 
However, as shown in Figure~\ref{fig:example}, each entity has at least one outgoing edge with a unique timestamp. As a result, following the 1-to-1 mapping assumption~\cite{MTransE17}, ``Japan'' can be aligned correctly, by simply hashing the entities using the timestamps of their outgoing edge. 
\vspace{-2mm}
\end{example}


We present an EA method for TKGs, \Method{}, that is able to address the above two problems. 
To tackle the first problem, instead of treating time intervals simply as relations~\cite{TEAGNN21, TREA22}, we model the temporal information using a learning-free encoder. Specifically, we derive pseudo seeds from the encoder to train GNN-based EA models. The models can encode relational information without any requirement for training data. 
To solve the second problem, we
develop a decoder that is independent of seed alignment. This is to fuse the temporal and relational information after they are encoded into feature vectors.
We transform EA into weighted graph matching, in order to wisely balance the temporal and relational information. Our contributions are summarized as follows:


\begin{itemize}[topsep=0pt,itemsep=0pt,parsep=0pt,partopsep=0pt,leftmargin=*]
    \item{\emph{Framework.}} We design \Method{}
    $\footnote{The codes are available at https://github.com/ZJU-DAILY/DualMatch/}$
    , an EA framework for TKGs, which effectively explores both temporal and relational information. 
    It contains two steps: (i) encoding the information from TKGs, and (ii) decoding the embedded data into an alignment matrix. 
    By employing an encoder \textbf{\Encoder{}} and a decoder \textbf{\Decoder{}}, \Method{} can perform EA on TKGs with or without training data.
    \item{\emph{Encoder.}} We develop a novel \Encoder{} that encodes the temporal and relational information separately. \Encoder{} has two components:
    (i) \TimeEncoder{} that retains the temporal information in TKGs; and
    (ii) \RelEncoder{} that learns the relational information using the graph structure.
    \item{\emph{Decoder.}} We design a novel \Decoder{} that balances the graph structure information and temporal information wisely, by building a bridge between the EA task and the weighted graph matching task. 
    \item{\emph{Experiments.}} We conduct comprehensive experiments on three real-world datasets, which suggest that \Method{} is able to outperform existing TKG \Revision{models}~\cite{TEAGNN21, TREA22} in terms of accuracy in both supervised and unsupervised settings.
\end{itemize}

\vspace*{-2mm}
\section{Related Work}
\label{sec:related_work}

\smallsection{Entity Alignment Encoders}
Most existing EA encoders are designed to learn the graph structures of KGs.
They can be divided into two categories: \emph{KGE}-\emph{based Encoder}~\cite{MTransE17, IPTransE17, BootEA18, AttrE19} and \emph{GNN-based Encoder}~\cite{GCN-Align18, KECG19, MRAEA20, AliNet20, HyperKA20, MuGNN19, DualAMN21}.
The former \Revision{uses} the KG embedding models (e.g., TransE~\cite{TransE13}) to learn entity embeddings, while the latter uses GNNs~\cite{GCN17}.
Recently, GNN-based models have demonstrated their outstanding performance~\cite{DualAMN21}. This is contributed by the strong modeling capability of GNNs to capture graph structure with anisotropic attention mechanism~\cite{GAT18}. However, they are not capable of modeling temporal information in TKGs.

\smallsection{Temporal EA Encoders}
\Revision{TEA-GNN~\cite{TEAGNN21} and TREA~\cite{TREA22} incorporate temporal information into the GNN architecture with timestamp.
However, these methods have not fully explored
the underlying efficacy of temporal information in GNN encoder embeddings and require seed alignment for EA.
On the contrary, 
the proposed \Encoder{} processes the temporal and relational information separately to improve the performance and is able to perform without seed alignment.
}

\smallsection{Entity Alignment Decoders}
Most existing EA decoders are designed to find equivalent entities based on the learned embeddings.
The most commonly used decoder is \emph{greedy search}~\cite{OpenEA2020VLDB}. It essentially finds the top-1-nearest neighbor using the similarity between embeddings of entities~\cite{MTransE17, IPTransE17, BootEA18, OpenEA2020VLDB}. To enhance the performance of EA, CSLS~\cite{CSLS} proposes to normalize the similarity matrix. 
However, the performance of CSLS' greedy top-1-nearest neighbor search is limited because it can only partly normalize the similarity matrix using local nearest neighbors.
CEA~\cite{CEAFF20} adopts the Deferred-acceptance algorithm (DAA)~\cite{GaleShapley} to find the stable matching between entities of two KGs, which produces higher-quality results than CSLS.
Nonetheless, DAA is hard to be parallelized on GPU as it performs iterations to match and un-match entity pairs. 
SEU~\cite{SEU21} transforms EA into the assignment problem by employing Hungarian algorithm~\cite{Hungarian1955} or Sinkhorn algorithm~\cite{Sinkhorn13} to normalize the similarity matrix. DATTI~\cite{DATTI22} decodes embeddings using Third-order Tensor Isomorphism. Recent studies ~\cite{LargeEA22, ClusterEA, LIME22, LargeScaleEACIKM1, LargeScaleEACIKM2} also present several decoders that exploit sampling to process large-scale EA. Different from previous decoders, our proposed \Decoder{} is designed for decoding features of TKGs. It needs two sets of input features, and effectively fuses them to generate high-quality alignment.

\smallsection{Unsupervised EA}
All existing proposals of unsupervised EA employ \emph{side information} of KGs, including
\emph{literal information} of entity names \cite{EASY21, SEU21} and \emph{visual information} \cite{EVA20}. Such proposals are able to perform EA without seed alignment ~\cite{OpenEA2020VLDB}.
However, models that incorporate machine translation or pre-aligned word embeddings may be overestimated due to the name bias issue~\cite{JEANS20, AttrGNN20, EVA20, NoMatch21}. In addition, it is difficult to extract visual information from KGs. On the contrary, our presented \Method{} only requires the information that is easily obtained from the TKGs in this paper and does not suffer from the name bias issue.


\vspace{-2mm}
\section{Preliminaries}
\label{sec:define}
We proceed to introduce preliminary definitions. Based on these, we formalize the problem of time-aware entity alignment.

\vspace{-2mm}
\subsection{Assignment Problem}

Assignment problem is a well-established combinatorial optimization problem.
It aims to find the optimal assignment $\boldsymbol{P}$ that transfers from a source distribution to a target distribution, while maximizing the overall profit.
Finding optimal $\boldsymbol{P}$ can be viewed as a special case of the optimal transport problem~\cite{Sinkhorn18}. 
 In this paper, we adopt the Sinkhorn algorithm~\cite{Sinkhorn13} to solve the assignment problem, which iterates $k$ steps for scaling the similarity/cost matrix.



An extension of the assignment problem is the quadratic assignment problem, which extends the assignment problem by defining the profit function in terms of quadratic inequalities. It is also commonly known as the 
weighted graph matching problem~\cite{GraphMatching96}.
Given two adjacency matrices $\boldsymbol{K_X} \in \mathbb{R}^{N \times N}$ and $\boldsymbol{K_Y} \in \mathbb{R}^{N \times N}$, the weighted graph matching problem aims to find an optimal  $\boldsymbol{P} \in \mathbb{P}_{N}$ such that $   \| \boldsymbol{K_X} \boldsymbol{P}-\boldsymbol{P} \boldsymbol{K_Y}\|_{2}^{2}$ is minimized.

\vspace{-2mm}
\subsection{Graph Kernel}

A graph kernel is a kernel function widely used in structure mining, which computes an inner product on graphs. Graph kernels can be used to compute how similar two graphs are. Thus, we use Weisfeiler-Lehman (WL) Subtree Kernel~\cite{WLKernels}, one of the most commonly-used graph kernels, to detect the similarity between two adjacency matrices. It calculates the similarity of graphs by running the WL Isomorphism Test~\cite{WLTest}. Specifically, given a set  $\mathcal{S}_G$ of graphs, we compute the WL kernel matrix $K_{WL}^{(h)} \in \mathbb{R}^{|\mathcal{S}_G|\times |\mathcal{S}_G|}$.
For each pair of graphs $(G_i, G_j)\in \mathcal{S}_G\times \mathcal{S}_G$, we have

\vspace{-2mm}
\begin{equation}
     k_{WL}^{(h)}(G_i, G_j) = \sum_{k = 0}^{h}\left\langle\phi\left(G_{i}^{(k)}\right), \phi\left(G_{j}^{(k)}\right)\right\rangle
\end{equation}
where $h$ is the number of WL iterations, $\left\{G^{(0)}, \ldots, G^{(h)}\right\}$ are the WL sequences of one graph $G$, and $\phi\left(G\right)$ indicates the feature mapping obtained from The WL Isomorphism Test. A traditional strategy to normalize the kernel matrix is to divide it by its diagonal, in which each element represents the similarity of a graph with itself, i.e., 
$
\hat{k}_{WL}(G_i, G_j) = \frac{ k_{WL}(G_i, G_j) }{\sqrt{diag(\boldsymbol{K}_{WL})\otimes diag(\boldsymbol{K}_{WL})}}.
$

\vspace{-2mm}
\subsection{Problem Definitions}
\begin{myDef}
A \textbf{temporal knowledge graph} (TKG) can be denoted as $G = (E,R,T, Q)$, where $E$ is the set of entities, $R$ is the set of relations, $T$ is the set of time intervals, and $Q=\{(h,r,t, \tau)~|~h,t \in E, r \in R, \tau \in T \}$ is the set of quadruples, each of which represents that the subject entity $h$ has the relation $r$ with the object entity $t$  during the time interval $\tau$. $\tau$ is represented as $[t_b, t_e]$,  with the beginning timestamp $t_b$ and the ending timestamp $t_e$.  
Some of the event-style quadruples, e.g., quadruples with the relation type ``born-in'',  may have $t_b = t_e$. 
\end{myDef}

\begin{myDef}
\label{sec:problem_statement}
\textbf{Time-aware Entity alignment} (TEA) \cite{TEAGNN21} aims to find a 1-to-1 mapping of entities $\phi$ from a source TKG\,$G_s = (E_s,R_s,T^*,Q_s)$ to a target TKG $G_t = (E_t,R_t,T^*, Q_t)$, where $T^*$ is a set of time intervals overlapped between
$G_s$ and $G_t$. $\phi = \{(e_s, e_t) \in E_s \times E_t~|~e_s \equiv e_t\}$ , where $e_s \in E_s$, $e_t \in E_t$, and $\equiv$ is an equivalence relation between two entities.
\end{myDef}

In \textbf{supervised} settings, a small set of equivalent entities $\phi^{\prime} \subset \phi$ is known beforehand, and is used as seed alignment. In \textbf{unsupervised} settings, the seed alignment $\phi^{\prime}$ is unknown.

\vspace{-2mm}
\section{Methodology}
We present our framework \Method{} for TKGs. We start by giving the overview of \Method{} and then detail each component of it.

\begin{figure*}[t]
\centering
\includegraphics[width=6.7in]{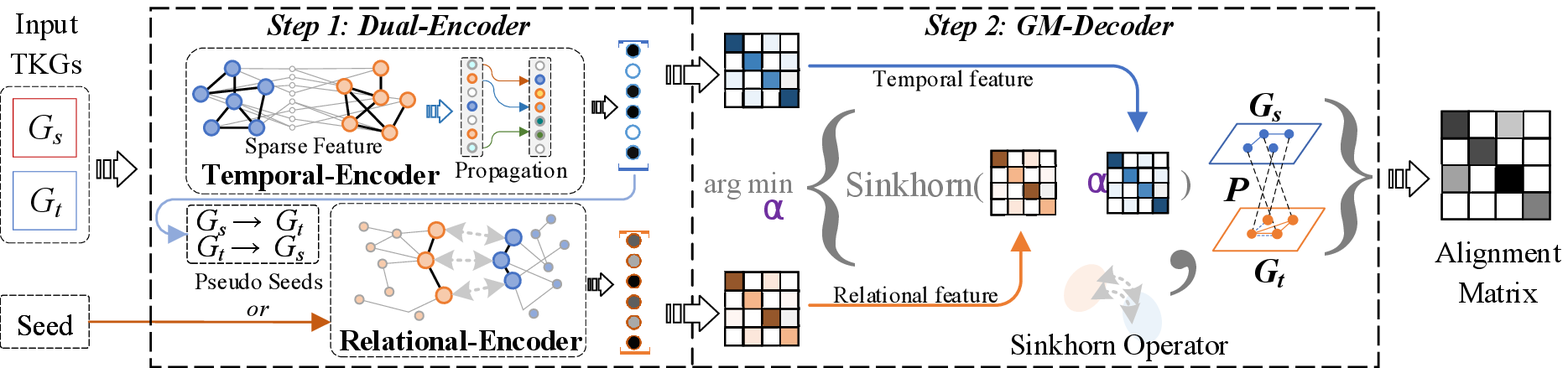}
\vspace*{-3mm}
\caption{\Method{} framework}
\label{fig:framewrok}
\vspace*{-4mm}
\end{figure*}

\vspace{-3mm}
\subsection{Overview}

Figure~\ref{fig:framewrok} shows \Method{} framework. The input (left side of Figure ~\ref{fig:framewrok}) includes two TKGs $G_s$,$G_t$, and an optional slot for seed alignment. In order to transfer the temporal and relational information of TKGs into a better alignment result, \Method{} employs a \Encoder{} (cf Step 1 of Figure~\ref{fig:framewrok}). It has two components to encode the input TKGs: 
(i) \TimeEncoder{} that aims at retaining the temporal information in TKGs; and
(ii) \RelEncoder{} that aims at learning the relational information using the graph structure.
With \Encoder{}, the temporal and relational information are encoded into separate entity features.
Next, \Method{} develops \Decoder{} (cf Step 1 of Figure~\ref{fig:framewrok}), which effectively fuses the temporal features with relational features by transforming the EA task into the weighted graph matching task. After the transformation, the \Decoder{} can produce a near-optimal fused alignment matrix (right side of Figure~\ref{fig:framewrok}).
Algorithm~\ref{alg:process} depicts the pseudo-code of the \Method{} framework.

\vspace{-2mm}
\subsection{Encoding Temporal Information}

Since the two TKGs to be aligned share the same set of time intervals that are recorded in quadruples, it is reasonable to make the following assumption: if two entities $e_s \in E_s$ and  $e_t \in E_t$ have many overlapped time intervals between their connected quadruples, $e_s$ and $e_t$ are very likely to represent the same real-world object.  
Based on this assumption, we propose a simple yet effective \emph{\TimeEncoder{}}.
 \TimeEncoder{} builds temporal features for entities in TKGs using their overlapped time intervals. For a TKG, we extract a bipartite graph $A^t$ between $E$ and $T$. 
 For each $a^t_{e t}\in A^t$, we have
 \begin{equation}
 \label{eq:time_adj}
 a^t_{e t} =  \frac{\text{exp}(|Q_{e t}|)}{\sum_{\tau \in T}\text{exp}(|Q_{e \tau}|)},
 \end{equation}
where $|Q_{e t}|$ is the number of quadruples that contain both entity $e$ and time interval $t$. Here, we separately calculate and concatenate the temporal adjacency matrices for head entities and tail entities. Thus, the result dimension of $A^t$ is $\left|E\right| \times\left|2T\right|$.
We call this bipartite graph the \textit{temporal feature matrix} of $E$. 

With the temporal feature matrices, $\boldsymbol{A}^t_s$ of $G_s$ and $\boldsymbol{A}^t_t$ of $G_t$, an alignment matrix can be easily derived by calculating the pairwise similarity: \Revision{$ \boldsymbol{P} = \boldsymbol{A}^t_s(\boldsymbol{A}^t_t)^T$}. However, some entities may not contain sufficient temporal information, which leads to poor alignment results. 
To solve this problem, \Revision{we impute the temporal feature from their neighborhoods, which may have a link to sufficient temporal facts, by aggregating information from neighbors; thus follow~\cite{SEU21} to }perform a $L$-layers graph-convolution-like forward pass to obtain an aggregated feature matrix: $\boldsymbol{H}^t = \left[\boldsymbol{A^t} ||  \boldsymbol{A} \boldsymbol{A^t}|| \boldsymbol{A}^{2} \boldsymbol{A^t}||...||\boldsymbol{A}^{L} \boldsymbol{A^t}\right]$ \CFLines{3}{4},
 where $L$ is a hyper-parameter indicating number of layers of the graph-convolution-like forward pass, $[X||Y]$ indicates tensor concatenate operations between $X$ and $Y$ along the last dimension,
 and $\boldsymbol{A}^{l} = \prod_{i \in 1,...,l}\boldsymbol{A}$ is the $l$-hop adjacency matrix of $G$. 
Here, to leverage the impact of different relation types on the knowledge graph, we follow~\cite{MRAEA20, RREA20, DualAMN21, SEU21, DATTI22} to build the \emph{relational feature matrix}. Specifically, for each $\boldsymbol{a}_{i j} \in \boldsymbol{A}$ , 
\begin{equation}
 \label{eq:rel_adj}
\boldsymbol{a}_{i j}=\frac{\sum_{r_{j} \in R_{i, j}} \ln \left(|Q| /\left|Q_{r_{j}}\right|\right)}{\sum_{k \in \mathcal{N}_{i}} \sum_{r_{k} \in R_{i, k}} \ln \left(|Q| /\left|Q_{r_{k}}\right|\right)},
\end{equation}
where $\mathcal{N}_{i}$ represents the neighboring set of entity $e_{i}, R_{i, j}$ is the relation set between $e_{i}$ and $e_{j}$, and $|Q|$ and $\left|Q_{r}\right|$ denote the total number of all quadruples and the quadruples containing relation $r$, respectively.

\vspace{-2.5mm}
\subsection{Encoding Relational Information}
State-of-the-art methods for encoding relational information typically use GNNs~\cite{GCN17} to propagate the information provided by the seed alignment $\phi^\prime$.  
Studies~\cite{TEAGNN21, TREA22} suggest that integrating temporal information can improve the performance of EA. They encode entities' relational features together with temporal features using a GNN. 
Specifically, they train the entity's embedding by propagating the neighborhood information~\cite{RREA20, GCN17, GAT18}.
The embedding of an entity $v \in E$ in the $l^{th}$ layer of GNN $h_v^{(l)}$ is computed by aggregating localized information:

\vspace{-4.5mm}
\begin{equation}
\begin{array}{c}
a_{v}^{(l)}=\operatorname{AGGREGATE}^{(l)}\left(\left\{\left(h_{u}^{(l-1)}, \textbf{r}_{u v},  \textbf{t}_{u v}\right)\mid u \in \mathcal{N}(v)\right\}\right) \\
h_{v}^{(l)}=\operatorname{UPDATE}^{(l)}\left(a_{v}^{(l)}, h_{v}^{(l-1)}\right),
\end{array}
\label{eq:message_passing}
\vspace{-2mm}
\end{equation}

\noindent where $h_v^0 \in \mathbb{R}^D$ is a learnable embedding vector initialized using Glorot initialization, $\textbf{r}_{u v} \in R$,  $\textbf{t}_{u v} \in T$ represents the relation types and time intervals between entity $u$ and $v$, and 
$\mathcal{N}(v)$ is the set of neighboring entities around $v$. The model's final output is denoted as $\boldsymbol{H}^{r} \in \mathbb{R}^{|E_s \cup E_t|\times D}$, where $D$ is the dimension of embeddings. 
 
\RelEncoder{} adopts the state-of-the-art EA method~\cite{DualAMN21} to build a $L$-layer GNN model for learning the representation of the knowledge graph structure \CFLine{9}. To be more specific, the dual-aspect embedding in~\cite{DualAMN21} concatenates both entity embeddings and relation embeddings. We extend it with additional time embeddings to form an overall \textit{triple-aspect embedding}. For each $\boldsymbol{h}_{e_{i}} \in \boldsymbol{H}^{r}$:
\vspace{-2mm}
\begin{equation}
    \boldsymbol{h}_{e_{i}}=\left[\boldsymbol{h}_{e_{i}}^{o u t} 
    \| \left(\frac{1}{\left|\mathcal{N}_{e_{i}}^{r}\right|} \sum_{r_{j} \in \mathcal{N}_{e_{i}}^{r}} \boldsymbol{h}_{r_{j}}
    + \frac{1}{\left|\mathcal{N}_{e_{i}}^{t}\right|} \sum_{t_{j} \in \mathcal{N}_{e_{i}}^{t}} \boldsymbol{h}_{t_{j}}\right)
    \right],
\end{equation}

\noindent where $\boldsymbol{h}_{e_{i}}^{o u t} = [h_{e_i}^{(0)}||h_{e_i}^{(1)}||...||h_{e_i}^{(L)}], \boldsymbol{h}_{r_{j}} $, and $ \boldsymbol{h}_{t_{j}} $ denote the vector of corresponding entity, relation, and time interval respectively, and  $\mathcal{N}_{e_{i}}^{r}$ and $\mathcal{N}_{e_{i}}^{t}$ represent the set of the relations and time intervals around entity $e_{i}$. \Revision{Note that recent works generally concatenate all the output features in each GNN layer~\cite{MRAEA20, RREA20, DualAMN21}, in order to achieve better performance. 
} Previous studies use negative sampling~\cite{MTransE17, AttrGNN20, MRAEA20} to form the loss function and to train the encoder model. We adopt this strategy for fast convergence~\cite{DualAMN21, TREA22}.
 The training loss is defined as $ \mathcal{L} = \text{LSE}(\lambda z(e_s, e_t)) + \text{LSE}(\lambda z(e_t, e_s))$
, where $e_s \in E_s$, $e_t \in E_t$, $(e_s, e_t) \in \phi^{\prime}$,
$ \text{LSE}(X) = log(\sum_{x \in X}e^x)$
is an operator to smoothly generates hard negative samples,
$\lambda$ is the smooth factor of $\text{LSE}$, and 
$z \in\mathcal{R}^{|E_t|}$ is the normalized triple loss. More specifically, $z$ is defined as $z(e_{s}, e_{t}) =\operatorname{z-score}(\{\gamma+\operatorname{sim}(e_{s}, e_{t})-\operatorname{sim}(e_{s}, e_{t}^{\prime}) | e^{\prime}_t \in E_t\}),$
in which $\operatorname{z-score}(X)=  \frac{X - \mu(X)}{\sigma(X)}$ is the standard score normalization,
and $\operatorname{sim}(e_{s}, e_{t})= \boldsymbol{h}^{r}_{e_s} \cdot \boldsymbol{h}^{r}_{e_t}$ is the similarity of two entities obtained by the dot product.

\vspace{-2mm}
\subsection{Fusing Information and Decoding}

\paragraph{Naive Strategy}
Previous studies have developed several techniques to improve EA performance by decoding embeddings into an alignment matrix. However, in \Decoder{}, the temporal and relational information is separately encoded into $\boldsymbol{H}^t$ and $\boldsymbol{H}^r$. 
A naive strategy is to directly concatenate two features by
$\boldsymbol{\hat{H}} = \left[ \boldsymbol{H}^{r} || \boldsymbol{H}^{t} \right]$. and then we can obtain the alignment matrix according to the concatenated features:
$
    \boldsymbol{P} = 
    \boldsymbol{\hat{H}}_{s}(\boldsymbol{\hat{H}}_{t})^T.
$

\paragraph{Our Strategy}
The naive decoding strategy has two problems. First, $\boldsymbol{H}^{r}$ is a dense matrix while $\boldsymbol{H}^{t}$ is sparse in practice.  Due to the commonly used format of storing sparse matrices, it is memory-consuming to concatenate one sparse matrix with a dense matrix. Second, $\boldsymbol{H}^{r}$ is obtained from Relational-Encoder while $\boldsymbol{H}^{t}$ is obtained from Temporal-Encoder. Thus they are not required to be considered equally. To address the aforementioned two problems, we introduce a weight $\alpha$: $    \boldsymbol{P} =\alpha\cdot
\boldsymbol{H}^{t}_{s}(\boldsymbol{H}_{t}^{t})^T 
+ \boldsymbol{H}^{r}_s(\boldsymbol{H}^{r}_t)^T$, 
\noindent where $\alpha$ is the weight to balance the impact of the two features on the alignment result.
Then, to normalize the alignment matrix, we adopt recent work~\cite{SEU21, DATTI22} that reformulated EA as an assignment problem and solved it via the Sinkhorn algorithm~\cite{SEU21, EASY21, ClusterEA, DATTI22}: $\boldsymbol{\hat{P}}= \text{Sinkhorn}\left(\boldsymbol{P}\right)$ \CFLine{13}. 
Note that, $|E_s|\not\eq|E_t|$ in many real-world scenarios. In this case, the EA problem is transformed into an unbalanced assignment problem, which can then be straightforwardly reformulated as a balanced assignment problem~\cite{SEU21}.

However,  the optimal $\alpha$ is hard to be determined because the ground truth is unknown. To solve this problem,
we study the isomorphic nature of the two TKGs to be aligned and propose to set the weight by minimizing the objective of weighted graph matching~\cite{GraphMatching96} with given $\boldsymbol{\hat{P}}$. More specifically, we map the source adjacency matrices into target ones. Next, by calculating the distance of the mapped matrix with the target matrix, we derive the correctness of the alignment matrix $\boldsymbol{\hat{P}}$ \CFLine{14}. 
\begin{equation}
\label{eq:gmd}
    \boldsymbol{D}_{G_s G_t} = k^r \left\|\boldsymbol{A}_{s}\boldsymbol{\hat{P}} - \boldsymbol{\hat{P}} \boldsymbol{A}_{t}\right\|_{2}^{2} + 
      k^t\left\|\boldsymbol{A}_{s}^{t} -\boldsymbol{\hat{P}} \boldsymbol{A}_{t}^{t}\right\|_{2}^{2},
\end{equation}

\noindent where $ k^r = \hat{k}_{WL}(\boldsymbol{A_s},\boldsymbol{A_t})$ and $ k^t = \hat{k}_{WL}(\boldsymbol{A_s^t},\boldsymbol{A_t^t})$ are the weights based on the isomorphism of the adjacency matrices.

Let $ d^r= \left\|\boldsymbol{A}_{s}\boldsymbol{\hat{P}} - \boldsymbol{\hat{P}} \boldsymbol{A}_{t}\right\|_{2}^{2}$ and $d^t=  \left\|\boldsymbol{A}_{s}^{t} -\boldsymbol{\hat{P}} \boldsymbol{A}_{t}^{t}\right\|_{2}^{2}$ be the \emph{distances} of the two TKGs measured by the relational and temporal feature matrices. 
Since the two TKGs may be non-isomorphic, and the relational adjacency matrices and temporal adjacency matrices are constructed independently, $d^r$ and $d^t$ may scale differently. Hence, it is reasonable to assign different weights to them.  
We denote $sim(\cdot)$ as the similarity between two adjacency matrices. 
Intuitively, if we have $sim(\boldsymbol{A_s}, \boldsymbol{A_t}) > sim(\boldsymbol{A^t_s}, \boldsymbol{A^t_t})$, $d^r$ can be weighted more. 
This motivates us to set the weights using the similarity of graphs measured by the WL Graph Kernels~\cite{WLKernels}.

\newcommand{\Comment}[1]{\textcolor{brown}{\tcc*[h]{#1}}}
\newcommand{\CommentNofill}[1]{ \textcolor{blue}{\tcc{#1}}}
\begin{algorithm}[t]
\small
\DontPrintSemicolon
  \SetKwProg{Fn}{def}{:}{}
  \KwInput{TKGs $G_s$, $G_t$, optional seed alignment $\phi^{\prime}$, and range $R_\alpha$ 
  }
  \KwOutput{Alignment matrix $\boldsymbol{\hat{P}}$ }
  
    \SetKwFunction{FTime}{\TimeEncoder{}}
    \SetKwFunction{FRel}{\RelEncoder{}}
    \SetKwFunction{FDecoder}{\Decoder{}}
\CommentNofill{Step $1$: encode temporal and relational information into vectors with \Encoder{}}
  $\boldsymbol{A}_s, \boldsymbol{A}^t_s,\leftarrow  G_s$,  $ \boldsymbol{A}_t, \boldsymbol{A}^t_t \leftarrow G_t$ \Comment{Extract  feature matrices}\;
  $\boldsymbol{H}^t_s\leftarrow$ \FTime{$\boldsymbol{A}_s, \boldsymbol{A}^t_s$} \;
  $\boldsymbol{H}^t_t \leftarrow$ \FTime{$\boldsymbol{A}_t, \boldsymbol{A}^t_t$}\;
  \If(\Comment{Compute pseudo-seeds}){$\phi^\prime$ is $not$ $given$}{
        $\boldsymbol{P}^{G_s G_t} \leftarrow \text{Sinkhorn}(\boldsymbol{H}^t_s(\boldsymbol{H}^t_t)^T)$ \;
        $\boldsymbol{P}^{G_t  G_s}\leftarrow \text{Sinkhorn}(\boldsymbol{H}^t_t(\boldsymbol{H}^t_s)^T)$\;
        $\phi^\prime\leftarrow \{(i,j) | \arg \max_{j^{\prime}}{\boldsymbol{P}^{G_s  G_t}_{ij^{\prime}}} = j  \wedge \arg \max_{i^{\prime}}{\boldsymbol{P}^{G_t  G_s}_{ji^{\prime}}} = i\}$ 
    }
  
  $\boldsymbol{H}^r_s, \boldsymbol{H}^r_t \leftarrow$ \FRel{$G_s, G_t, \phi^\prime$}\;
   \CommentNofill{Step $2$: decode feature vectors with \Decoder{}}
  $\boldsymbol{\hat{P}}, \boldsymbol{D} \leftarrow zeros(|E_s|, |E_t|), +\infty$\; 
  \For{$\alpha \in R_\alpha$}{
     $\boldsymbol{\hat{P}}_{\alpha} \leftarrow \text{Sinkhorn}(\alpha\cdot
    \boldsymbol{H}^{t}_{s}(\boldsymbol{H}_{t}^{t})^T 
    + \boldsymbol{H}^{r}_s(\boldsymbol{H}^{r}_t)^T)$\;
    $\boldsymbol{D}_{G_s G_t}^{\alpha} = k^r \left\|\boldsymbol{A}_{s}\boldsymbol{\hat{P}}_\alpha - \boldsymbol{\hat{P}}_\alpha \boldsymbol{A}_{t}\right\|_{2}^{2} + 
      k^t\left\|\boldsymbol{A}_{s}^{t} -\boldsymbol{\hat{P}}_\alpha \boldsymbol{A}_{t}^{t}\right\|_{2}^{2}$\;
    \If{$\boldsymbol{D}_{G_s G_t}^{\alpha} < \boldsymbol{D}$}{
        $\boldsymbol{D}, \boldsymbol{\hat{P}}\leftarrow\boldsymbol{D}_{G_s G_t}^{\alpha},  \boldsymbol{\hat{P}}_\alpha$\Comment{Update alignment matrix}
    }
  }
  
  \KwRet $\boldsymbol{\hat{P}}$. 
  
\caption{The proposed \Method{} process}
\label{alg:process}
\end{algorithm}


We search $\alpha=\arg\min_{\alpha} \boldsymbol{D}_{G_s G_t}$ in the range of $R_\alpha$, in order to obtain the optimal $\boldsymbol{\hat{P}}$ \CFLines{11}{16}. When matching two graphs using a mapping matrix, we do not need to map more than one entity from the source graph into the target graph.  Here $zeros(S)$ indicates a Tensor of shape $S$ filled with $0$. However, there are multiple nonzero values in each row of $\boldsymbol{\hat{P}}$. This results in that, for one entity in the source TKG, several possible alignments exist in the target TKG with weighted scores.
To this end, we decide to sparsify $\boldsymbol{\hat{P}}$ by only retaining the top-$1$ correspondence. To keep the matrix doubly scholastic, we set the value to 1 for the top-$1$ correspondence:

\vspace{-3mm}
\begin{equation}
\label{eq:p}
\boldsymbol{\hat{P}}_{ij}=\left\{\begin{array}{ll}
1, & j=\arg\max_{j'}(\boldsymbol{\hat{P}}_{ij'}) \\
0, & \text{otherwise}
\end{array}\right.
\end{equation}

\subsection{Unsupervised Learning}

Since the \RelEncoder{} needs seed alignment, in order to perform unsupervised EA, we follow the Bi-directional Strategy~\cite{MRAEA20} to generate pseudo-seeds from \TimeEncoder{}.
Considering the asymmetric nature of alignment directions~\cite{ LargeEA22, ClusterEA}, we first derive the bi-directional alignment matrices $\boldsymbol{P}^{G_s G_t} = \text{Sinkhorn}(\boldsymbol{H}^t_s(\boldsymbol{H}^t_t)^T)$ and  $\boldsymbol{P}^{G_t  G_s}= \text{Sinkhorn}(\boldsymbol{H}^t_t(\boldsymbol{H}^t_s)^T)$ from the temporal feature by applying \TimeEncoder{} on both directions \CFLines{6}{7}. This process can be performed without supervision. Then, we extract pseudo-seeds from the two matrices. Specifically, we consider $(e_i, e_j) \text{ where } e_i \in E_s, \; e_j \in E_t$ as a valid seed pair iff $\arg \max_{j^{\prime}}{\boldsymbol{P}^{G_s  G_t}_{ij^{\prime}}} = j $ and $\arg \max_{i^{\prime}}{\boldsymbol{P}^{G_t  G_s}_{ji^{\prime}}} = i$ \CFLine{8}.  Finally, with the generated pseudo-seed alignment, we train \RelEncoder{} \CFLine{9} and obtain the unsupervised alignment result $\boldsymbol{\hat{P}}$ via the \Decoder{} \CFLines{11}{17}.  

\smallsection{Discussion}
Many previous studies have tried to generate pseudo-seeds from machine translation systems or pre-trained embeddings~\cite{JAPE17, EVA20, MRAEA20, RREA20, DualAMN21} to gain better EA performance.
Unlike them, we generate the pseudo-seed with the information that is an inherent part of TKGs. This means that \Method{} can fully exploit the inherent information of datasets to perform EA rather than relying on information that is hard to be obtained.



\vspace{-2mm}
\section{Experiments}
\label{sec:exp}

\begin{table*}[t]\small
\caption{Overall EA results on \DICEWS{} and \WIKIYAGO{}
}\label{exp:overall_15K_100K}
\vspace*{-4mm}
\begin{threeparttable}
\setlength{\tabcolsep}{1.5mm}{
\begin{tabular}{clccclccclccclccc}
\toprule 
\multicolumn{2}{c}{\multirow{2}{*}{\textbf{Methods}}} &
  \multicolumn{3}{c}{\DICEWS{-1K}} &
   &
  \multicolumn{3}{c}{\DICEWS{-200}} &
   &
  \multicolumn{3}{c}{\WIKIYAGO{-5K}} &
   &
  \multicolumn{3}{c}{\WIKIYAGO{-1K}} \\ \cline{3-5} \cline{7-9} \cline{11-13} \cline{15-17} 
\multicolumn{2}{c}{}& H@1  & H@10 & MRR   &  & H@1  & H@10 & MRR   &  & H@1  & H@10 & MRR   &  & H@1  & H@10 & MRR   \\ \hline
\multirow{10}{*}{Time-Unaware} & MTransE    & 10.1 & 24.1 & 0.150  &  & 6.7  & 17.5 & 0.104 &  & 24.2 & 47.7 & 0.322 &  & 1.2  & 6.7  & 0.033 \\
  & JAPE  & 14.4 & 29.8 & 0.198 &  & 9.8  & 21.0   & 0.138 &  & 27.1 & 48.8 & 0.345 &  & 10.1 & 26.2 & 0.157 \\
  & AlignE& 50.8 & 75.1 & 0.593 &  & 22.2 & 45.7 & 0.303 &  & 75.6 & 88.3 & 0.800   &  & 56.5 & 71.4 & 0.618 \\
  & GCN-Align  & 20.4 & 46.6 & 0.291 &  & 16.5 & 36.3 & 0.231 &  & 51.2 & 71.1 & 0.581 &  & 21.7 & 39.8 & 0.279 \\
  & MuGNN & 52.5 & 79.4 & 0.617 &  & 36.7 & 58.3 & 0.412 &  & 76.2 & 89.0   & 0.808 &  & 58.9 & 73.3 & 0.632 \\
  & MRAEA & 67.5 & 87.0   & 0.745 &  & 47.6 & 73.3 & 0.564 &  & 80.6 & 91.3 & 0.848 &  & 62.3 & 80.1 & 0.685 \\
  & HyperKA    & 58.8 & 84.2 & 0.669 &  & 38.3 & 65.3 & 0.474 &  & 78.4 & 90.0   & 0.829 &  & 61.0   & 77.5 & 0.665 \\
  & RREA  & 72.2 & 88.3 & 0.780 &  & 65.9 & 82.4 & 0.719 &  & 82.8 & 93.8 & 0.868 &  & 69.6 & 85.9 & 0.753 \\
  & KE-GCN& 54.9 & 82.7 & 0.650  &  & 37.3 & 62.5 & 0.451 &  & 78.0   & 91.0   & 0.831 &  & 60.0   & 76.1 & 0.654 \\
  & Dual-AMN & 71.6 & 89.3 & 0.779 & &		66.8 &	85.4&	0.733& & 		89.7&	96.4&	0.922	& &	75.5 &	89.0& 	0.834\\ \hline
\multirow{3}{*}{Time-Aware}    & TEA-GNN    & 88.7 & 94.7 & 0.911 &  & 87.6 & 94.1 & 0.902 &  & 87.9 & 96.1 & 0.909 &  & 72.3 & 87.1 & 0.775 \\
  & TREA  & 91.4 & 96.6 & 0.933 &  & 91.0   & 96.0   & 0.927 &  & 94.0   & 98.9 & 0.958 &  & 84.0   & 93.7 & 0.885 \\
  & \Method{}  & \textbf{95.3} & \textbf{97.3} & \textbf{0.961} &  & \textbf{95.3} & \textbf{97.4} & \textbf{0.961} &  & \textbf{98.1} & \textbf{99.6} & \textbf{0.986} &  & \textbf{94.7} & \textbf{98.4} & \textbf{0.961} \\
  \hline
\multirow{1}{*}{Unsupervised}  & \Method{} (unsup)  & \textbf{94.6} & \textbf{97.1} & \textbf{0.956} &  & -    & -    & -&  & \textbf{96.4} & \textbf{99.1} & \textbf{0.975} &  & -    & -    & -\\
\bottomrule
\end{tabular}

}
\end{threeparttable}
\end{table*}

We report on extensive experiments aiming at evaluating the performance of \Method{}.
\vspace{-3mm}
\subsection{Experimental Setup} 
\label{sec:exp_setup}
\smallsection{Datasets}
We use three real-world TKG datasets provided by~\cite{TEAGNN21, TREA22}. They are extracted from ICEWS05-15~\cite{ICEWS05-15}, YAGO~\cite{YAGO}, and Wikidata~\cite{Wikidata14}, denoted as \DICEWS{}, \WIKIYAGO{} and \WIKIYAGOHybrid{}, respectively. The detailed statistics of all datasets are presented in Appendix~\ref{app:dataset_stat}.

\smallsection{Baselines}
We use 12 state-of-the-art EA methods as baselines. We follow the settings in~\cite{EVA20, RREA20, AttrGNN20, DualAMN21, ClusterEA, TREA22, TEAGNN21}, which exclude side information other than the temporal and relational information in TKGs. This is to guarantee a fair comparison.
We divide the baselines into two categories below. 

\vspace{4pt}
\begin{itemize}[topsep=0pt,itemsep=0pt,parsep=0pt,partopsep=0pt,leftmargin=*]
    \item \textbf{Time-Unaware baselines} that do not exploit the temporal information to perform EA, including 
        (1) \emph{MTransE}~\cite{MTransE17}, which is the first embedding-based EA model; 
        (2) \emph{JAPE}~\cite{JAPE17}, which employs attribute correlations for entity alignment;
        (3) \emph{AlignE}~\cite{BootEA18}, which trains KG embedding in  an  alignment-oriented fashion;
        (4) \emph{GCN-Align} \cite{GCN-Align18}, which aligns entities via graph convolutional networks;
        (5) \emph{MuGNN} \cite{MuGNN19}, which learns alignment-oriented embeddings by a multi-channel graph neural network;
        (6) \emph{MRAEA} \cite{MRAEA20}, which learns cross-graph entity embeddings with the entity’s neighbors and its connected relations’ meta semantics; 
        (7) \emph{HyperKA}~\cite{HyperKA20}, which learns the hyperbolic embeddings for EA;
        (8) \emph{RREA}~\cite{RREA20}, which  leverages relational reflection transformation to obtain relation-specific embeddings for EA;
        (9) \emph{KE-GCN}~\cite{KEGCN21}, which learns a knowledge-embedding based graph convolutional network; and
        (10) \emph{Dual-AMN}~\cite{DualAMN21}, the state-of-the-art EA model, only uses the relational information of knowledge graph structure.
    \item \textbf{Time-Aware baselines} that consider both relational and temporal information, including
        (1) \emph{TEA-GNN}~\cite{TEAGNN21}, which models temporal information as embeddings; and 
        (2) \emph{TREA}~\cite{TREA22},  which is the state-of-the-art time-aware EA model using a temporal relation attention mechanism.
\end{itemize}
\vspace{4pt}

\smallsection{Evaluation Metrics}
The widely-adopted \emph{\HitNFull{} (\HitN{})} $(N=1, 10)$ and \emph{Mean Reciprocal Rank (MRR)} are used as the evaluation metrics~\cite{MTransE17, BootEA18,  DualAMN21, DATTI22}.
\HitNFull{} (in percentage) denotes the proportion of correctly aligned entities in the top-$N$ ranks in the alignment matrix $\boldsymbol{\hat{P}}$.
MRR is the average of the reciprocal ranks of the correctly aligned entities, where reciprocal rank reports the mean rank of the correct alignment derived from $\boldsymbol{\hat{P}}$.
Note that, higher \HitN{} and MRR indicate higher EA accuracy.

\smallsection{Experimental Settings}
In order to avoid the labor-intensive seed alignment, we expect to use as less training data as possible.
Traditional EA methods typically use 20\%-30\% of the total number of pairs to train the EA model
~\cite{OpenEA2020VLDB}. The temporal information can be treated as an alternative for seed alignment of TKGs, and thus, fewer training pairs need to be used.
We use three \emph{settings} of train pair ratio to evaluate the EA performance, including two \emph{supervised} settings that follow previous studies~\cite{TEAGNN21, TREA22}, and an \emph{unsupervised} setting that uses no seeds.
\begin{itemize}[topsep=0pt,itemsep=0pt,parsep=0pt,partopsep=0pt,leftmargin=*]
    \item \textbf{Normal setting} uses roughly 10\% seeds for \DICEWS{}  and \WIKIYAGO{}, 
    which contain 1000 pairs and 5000 pairs, respectively, denoted as \DICEWS{}-1K and \WIKIYAGO{}-5K. The rest of the pairs are utilized to verify the EA performance.
    \item \textbf{Less seed setting} uses 200 and 1000 seeds for \DICEWS{} and \WIKIYAGO{}, respectively, denoted as \DICEWS{}-200 and \WIKIYAGO{}-1K. The rest of the pairs are used to evaluate the EA performance.
    \item \textbf{Unsupervised setting}\footnote{The unsupervised setting uses all the alignment pairs to evaluate. For a fair comparison, we compare it with the normal settings of baseline models.} does not contain seed alignment, and all the entity pairs are used for evaluation. As the previous methods require labeled seed alignments, we only provide the unsupervised EA results of \Method{}, denoted as \Method{} (unsup).
\end{itemize}

\noindent
We implement baselines by following their original settings.
Specifically, \textbf{Bold} digits in Tables 2--5 indicate the highest performance of supervised and unsupervised settings.
In \Method{}, for \Encoder{}, 
we follow previous studies~\cite{DualAMN21, SEU21} to set the hyper-parameters for aggregating temporal features and training the GNN-based EA model; for \Decoder{}, we search the best $\alpha$ minimizing $\boldsymbol{D}_{G_sG_t}$ in  $R_\alpha=\{0,$ $0.25,$ $0.5,$ $0.75,$ $1,$ $1.25, 1.5,$ $1.75,$ $2\}$.We set the number of WL iteration $h= 8$, and the step of Sinkhorn $k=10$ following~\cite{SEU21}.
All algorithms are implemented in Python, and the experiments are run on a computer with an Intel Core i9-10900K CPU, an NVIDIA GeForce RTX3090 GPU and 128GB memory. The artifact is available at \textit{https://doi.org/10.5281/zenodo.7149372}. 

\vspace{-2mm}
\subsection{Comparison}

\smallsection{Supervised Performance}\label{sec:exp_ids}
Table~\ref{exp:overall_15K_100K} summarizes the EA performance in supervised settings on \DICEWS{} and \WIKIYAGO{}.
First, \Method{} achieves the best performance in terms of all metrics on both datasets. \Method{}  improves \HitOne{} by $3.9\%-10.7\%$ compared with two state-of-the-art time-aware baselines, i.e., TEA-GNN~\cite{TEAGNN21} and TREA~\cite{TREA22}. This validates the superiority of the way we incorporate temporal information. Specifically, TEA-GNN~\cite{TEAGNN21} and TREA~\cite{TREA22} simply model the temporal information as learnable weight matrices, while \Method{} captures the temporal information in TKGs, which enables EA to use a separated \TimeEncoder{}. 
Second, all time-aware methods perform better than the time-unaware methods in terms of \HitOne{}. It confirms the superiority of incorporating temporal information in EA. Third, \Method{} improves \HitOne{} by $8.4\%-28.5\%$ compared to the time-unaware methods, which demonstrates the accuracy of \Method{}. 
Finally, among all the time-unaware baselines, Dual-AMN~\cite{DualAMN21} achieves the best alignment performance. The reason is that the design of its GNN model best captures the relational information hidden in the graph structure of TKGs. This implies the proposed \RelEncoder{}, which is extended from it, can learn the relational information accurately.
 
\smallsection{Unsupervised Performance}
By employing a bi-directional strategy, \TimeEncoder{} generates 8168, 35403, and 9778 pseudo seed pairs for \DICEWS{}, \WIKIYAGO{} and \WIKIYAGOHybrid{}, respectively. Next, we train \RelEncoder{} based on the pseudo seeds and report the performance of the final alignment matrix in unsupervised settings in
Table~\ref{exp:overall_15K_100K}. 
It can be observed that \Method{} outperforms TREA, which is the best baseline, by $2.4\%-3.2\%$ in terms of  \HitOne{}. 
Moreover, \Method{} improves \HitOne{} by $3.6\%-12.4\%$ compared with TREA~\cite{TREA22}, which achieves the best performance among all baselines in fewer seed settings. 
Overall, \Method{} is able to achieve better EA in both supervised and unsupervised settings than SOTAs. 

\subsection{Ablation Study}

\newcommand{\Remove}[1]{\Method{} - #1}

\begin{table}[t]\small
\caption{The result of ablation study}\label{exp:ablation}
\vspace*{-4mm}
\setlength{\tabcolsep}{1.mm}{
\begin{tabular}{lccccccc}
\toprule
\multicolumn{1}{c}{\multirow{2}{*}{\textbf{Method}}} & \multicolumn{3}{c}{\DICEWS{}-200} &  & \multicolumn{3}{c}{\WIKIYAGO{}-1K} \\ \cline{2-4} \cline{6-8} 
\multicolumn{1}{c}{} & H@1  & H@10 & MRR   &  & H@1  & H@10 & MRR   \\ \hline
\Method{}                 & \textbf{95.3} & \textbf{97.4} & 0.961 &  & \textbf{94.7} & \textbf{98.4} & \textbf{0.961} \\
 \Remove{TE}                & 90.3 & 95.2 & 0.922 &  & 88.4 & 96.2 & 0.913 \\
 \Remove{RE}                & 92.8 & 97.0 & 0.944 &  & 67.8 & 79.7 & 0.720 \\
 \Remove{TAL}               & 91.9 & 95.8 & 0.934 &  & 91.9 & 96.6 & 0.936 \\
 \Remove{TA}                & 94.1 & 96.5 & 0.951 &  & 94.2 & 98.1 & 0.957 \\
 \Remove{$\alpha$}              & \textbf{95.3} & \textbf{97.4} & \textbf{0.962} &  & 93.9 & 98.2 & 0.955 \\
 \Remove{GM-D}          & 91.3 & 94.8 & 0.927 &  & 80.4 & 92.2 & 0.845 \\
\bottomrule
\end{tabular}
}
\vspace*{-4mm}
\end{table}

We remove
each component of \Method{}, and report \HitOne{}, \HitTen{}, and \MRR{} in Table~\ref{exp:ablation}.
First, after removing the \TimeEncoder{} component (\Remove{TE}), the accuracy drops. This suggests that the \TimeEncoder{} indeed captures more temporal information.
Second, after removing the \RelEncoder{} component (\Remove{RE}), \HitOne{} of \Method{} decreases. This confirms the effectiveness of \RelEncoder{} in encoding the relational information.  Note that  removing \TimeEncoder{} has a similar impact on accuracy with removing \RelEncoder{} on \DICEWS{}; while on \WIKIYAGO{}, removing \RelEncoder{} incurs a significant drop in accuracy. This is because the importance of temporal and relational information varies from different datasets. Specifically, the relational information is more important for achieving better performance on \WIKIYAGO{}.
Third, by simply setting $\alpha=1$ (\Remove{$\alpha$}), the accuracy of all methods on \DICEWS{} remains almost the same but drops on \WIKIYAGO{}. This also shows that, in \DICEWS{}, temporal information is relatively as important as relational information, but in \WIKIYAGO{}, relational information is much more important than temporal information. More details about it will be covered in Section~\ref{sec:decoder}.
Fourth, by removing the graph-convolutional-like aggregation in \TimeEncoder{} (\Remove{TA}), the accuracy drops. This implies that the aggregation process enables the propagation of temporal information to improve total performance.
Next, by replacing the triple-aspect-learning with the original dual-aspect-learning (\Remove{TAL}), the accuracy drops. This suggests that incorporating temporal information into the learning process is the key to improving the effectiveness of EA. 
Finally, by replacing \Decoder{} with the \emph{Naive Strategy} described in  Section~\ref{sec:decoder} (\Remove{GM-D}), the accuracy of \Method{} drops significantly. This validates the effectiveness of \Decoder{}.

\subsection{Sensitivity Study}
\label{sec:sensitive}

\begin{table}[t]\small
\caption{The result of sensitive analysis}\label{exp:sensitive}
\vspace*{-4mm}
\setlength{\tabcolsep}{1mm}{
\begin{tabular}{lccclccc}
\toprule
\multicolumn{1}{c}{\multirow{2}{*}{\textbf{Method}}} & \multicolumn{3}{c}{Highly Time-Sensitive} &  & \multicolumn{3}{c}{Lowly Time-Sensitive} \\ \cline{2-4} \cline{6-8} 
\multicolumn{1}{c}{} & H@1  & H@10  & MRR   &  & H@1  & H@10 & MRR   \\ \hline
\Method{}                 & 99.1 & \textbf{100.0} & 0.994 &  & 35.9 & 49.7 & 0.407 \\
\Method{} (unsup)          & \textbf{99.8} & \textbf{100.0} & \textbf{0.999} &  & \textbf{43.1}	&\textbf{52.4}	 &\textbf{0.467}     \\
TREA                 & 87.5 & 96.1  & 0.909 &  & 31.9 & 47.0 & 0.370  \\
TEA-GNN              & 85.3 & 95.0  & 0.888 &  & 31.9 & 44.9 & 0.364
\\\bottomrule
\end{tabular}
}
\end{table}

In real-world applications, certain entities are not affected by temporal information.
Thus, we perform a study of the alignment result's sensitivity to the temporal information, using the temporal-hybrid dataset \WIKIYAGOHybrid{} that contains certain non-temporal facts. 
Following~\cite{TEAGNN21, TREA22}, we divide entity pairs into \emph{highly time-sensitive} entity pairs and \emph{lowly time-sensitive} entity pairs according to the ratio of the number of entities' connected temporal facts over the amount of all facts connected. 400 pairs are used as seeds for training the supervised models, 6,898 are highly time-sensitive pairs, and the rest is lowly time-sensitive pairs.  

We compare the EA performance of \Method{} with the time-aware baselines, and the result is reported in Table~\ref{exp:sensitive}. We also report the result of \Method{} by performing unsupervised EA on the dataset, where the 400 seeds are discarded. 
First, we observe that \Method{} outperforms previous methods significantly in terms of aligning both highly sensitive pairs and lowly sensitive pairs. This shows the overall effectiveness of \Method{} on real-world applications where the temporal information is insufficient.
Moreover, \HitOne{} of \Method{} is almost 100\% for the highly time-sensitive pairs. This confirms that \Method{} can capture the temporal information maximally with the separated \TimeEncoder{}. 
Second, on the lowly sensitive pairs,  \HitOne{} of \Method{} for two settings is higher than baselines, which validates that our overall design is more effective even for the entities with less temporal information. Third, the unsupervised variant of \Method{} dramatically outperforms the supervised one on the lowly sensitive pairs. This is because we utilize  the temporal-based pseudo seeds that could transfer temporal information into relational information. 
Finally, although TREA~\cite{TREA22} outperforms TEA-GNN~\cite{TEAGNN21} on both \DICEWS{} and \WIKIYAGO{} dataset, the performance of TREA is almost the same as TEA-GNN, especially on the lowly sensitive pairs. The reason is that the effect of temporal information is limited by the poor time embedding design on temporal-hybrid datasets~\cite{TREA22,TEAGNN21}. In contrast, \Method{}  employs the temporal and relational information in a better way and hence outperforms all the baselines.

\subsection{\Decoder{} Evaluation}
\label{sec:decoder}



\begin{figure}[t]
\centering
\includegraphics[width=.50\textwidth]{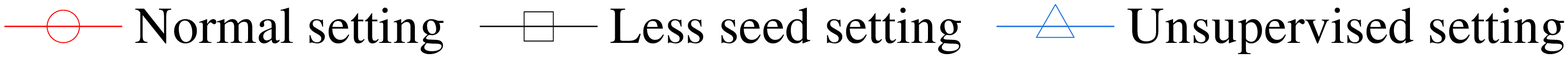}\\
\vspace*{-2mm}
\includegraphics[width=.45\textwidth]{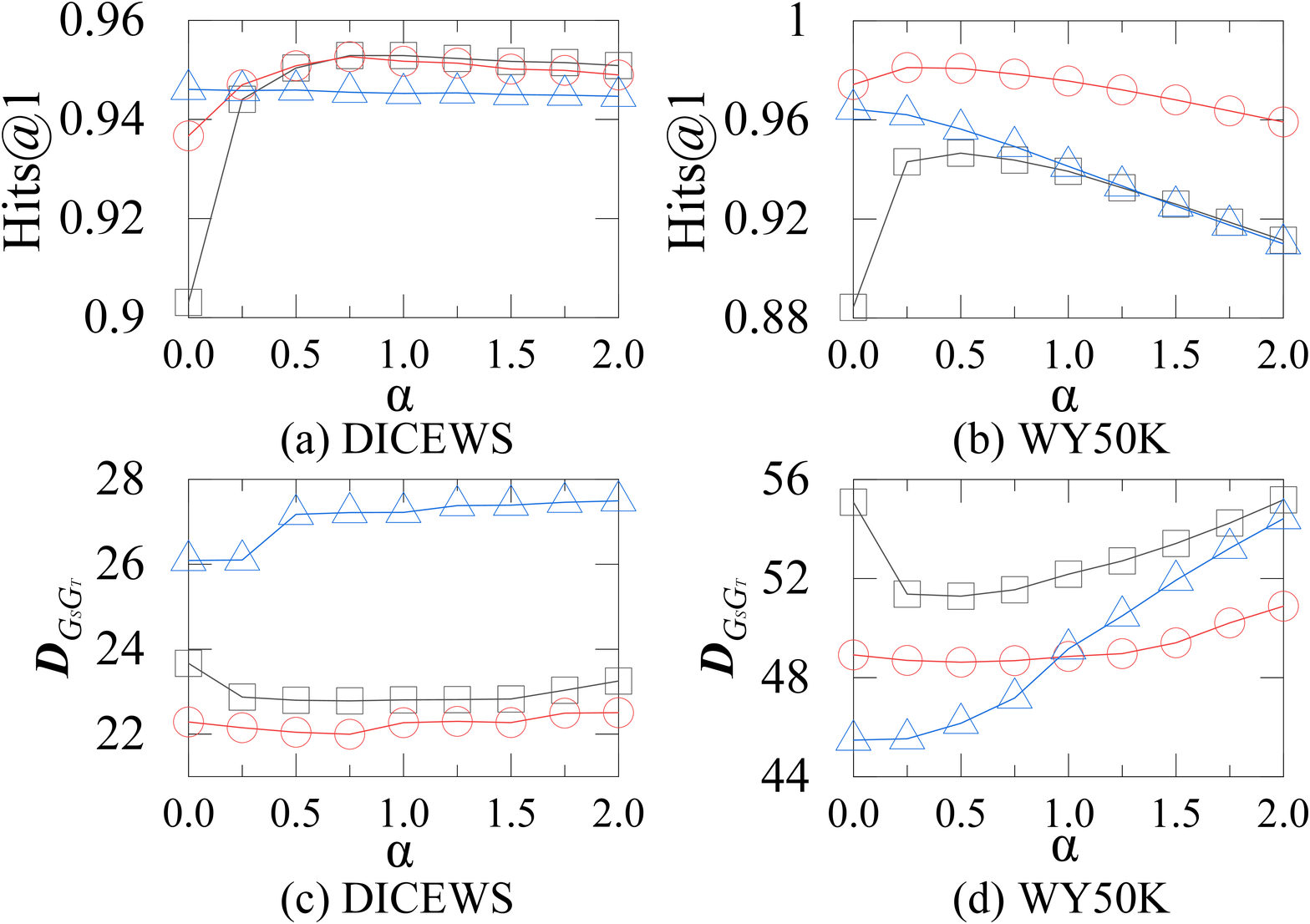}
\vspace*{-4mm}
\caption{The result of \Decoder{} analysis}
\vspace*{-4mm}
\label{fig:decoder-exp}
\end{figure}

\Decoder{} formulates an objective $\boldsymbol{D}_{G_sG_t}$ to optimize the EA performance by transforming the EA problem into the graph matching problem. 
Ideally, the accuracy should be inversely proportional to $\boldsymbol{D}_{G_sG_t}$, meaning that the lower $\boldsymbol{D}_{G_sG_t}$, the better the EA performance. This aligns with the objective of minimizing the distance.
To check whether it is the real situation, we proceed to evaluate the effectiveness of the \Decoder{} on finding optimal $\alpha$, by plotting the relation between the $\boldsymbol{D}_{G_sG_t}$ and the alignment performance.  
We vary $\alpha$ from 0 to 2 (cf. Section~\ref{sec:exp_setup}) in all three settings and
record \HitOne{} and $\boldsymbol{D}_{G_sG_t}$ of each step, as shown in Figure~\ref{fig:decoder-exp}. 
First, we observe that \HitOne{} is negatively correlated with $\boldsymbol{D}_{G_sG_t}$ for three settings, which validates the effectiveness of \Decoder{} in terms of finding the optimal $\alpha$. 
We also prove this in Table~\ref{exp:grid_search}, where we  report the worst and best \HitOne{} when varying $\alpha$. Table~\ref{exp:grid_search} shows that \Decoder{} finds the best $\alpha$ in all cases.
Second, the performance becomes better when $\alpha$ drops in the Unsupervised setting. This is because the pseudo-seed used in \RelEncoder{} is generated with \TimeEncoder{}, indicating that the temporal information is already learned by the \RelEncoder{}.
Third, the Normal setting generally outperforms the Less seed setting. This confirms that with more seed alignment, the \RelEncoder{} can capture more relational information.
Finally, we observe that the optimal $\alpha$ for \DICEWS{} is around $1$ while it is around $0.5$ for \WIKIYAGO{}. This implies that the importance of temporal and relational information in \DICEWS{} are almost the same, but that in \WIKIYAGO{} are biased. This may be caused by the algorithm generating \DICEWS{}~\cite{TEAGNN21}, which duplicates the ICEWS05-15 dataset into two TKGs with almost identical distributions. Based on the comparison results, \WIKIYAGO{} is fitter for real-world applications, where the amount of temporal and relational information is usually biased.

\vspace{-2mm}
\subsection{Scalability Study}
\label{exp:scalability}
\begin{figure}[t]
\centering
\includegraphics[width=0.46\textwidth]{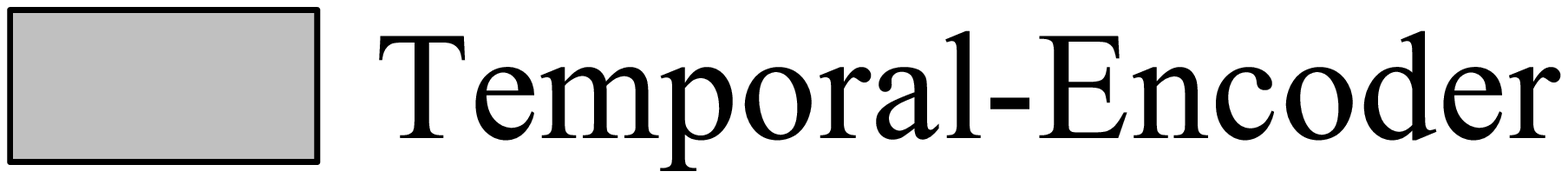}\vspace*{-1mm}\\
\includegraphics[width=0.46\textwidth]{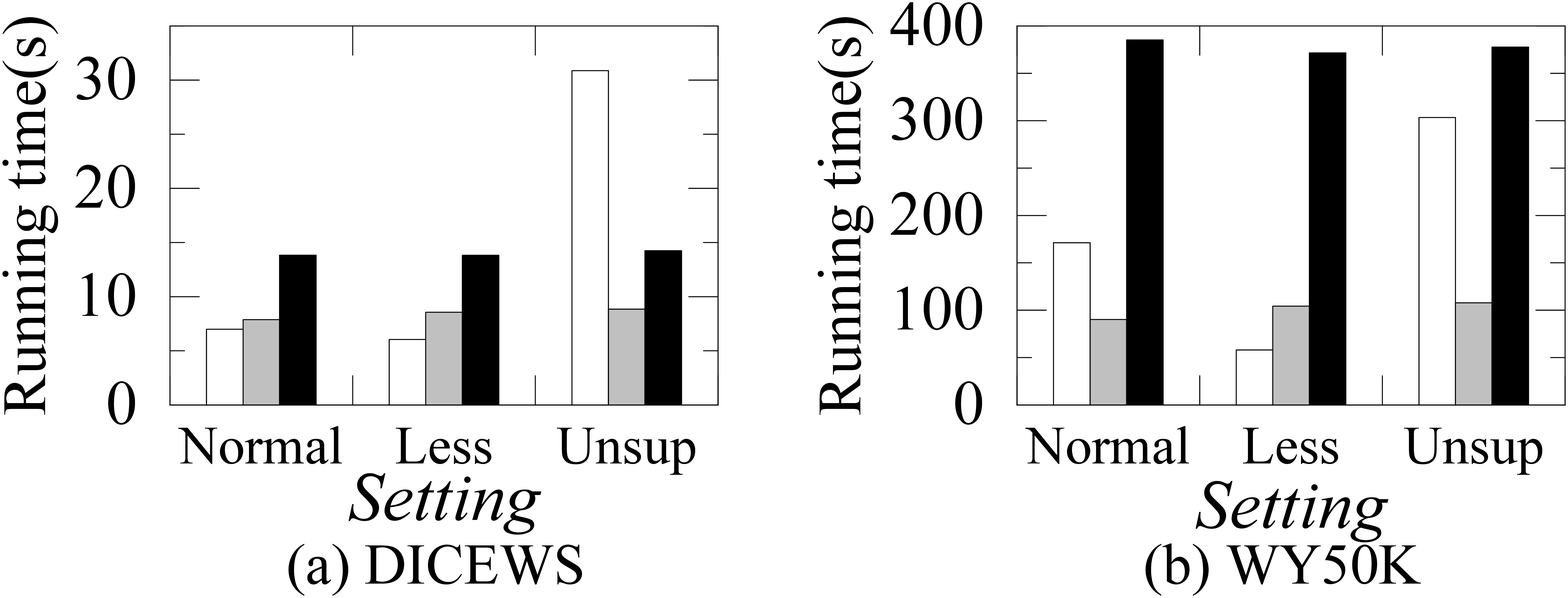}
\vspace*{-2mm}
\caption{Scalability analysis vs. settings}
\vspace*{-2mm}
\label{fig:scalability}
\end{figure}

To verify the scalability of \Method{}, we plot the running time of each component on \DICEWS{} and  \WIKIYAGO{} in Figure~\ref{fig:scalability}, where Normal, Less, and Unsup denote the Normal setting, Less seed setting, and Unsupervised setting, respectively. Specifically, the components are (1) \emph{\RelEncoder{}} that trains a GNN model to encode relational information; (2) \emph{\TimeEncoder{}} that takes the TKG inputs and encode the temporal information; and (3) \emph{\Decoder{}} that fuses the two encoded information to produce the alignment matrix.  
We observe that the running time taken by each component does not exceed $10^3$ seconds on both \DICEWS{} and \WIKIYAGO{}, which confirms the scalability of \Method{} on a large dataset. We also observe that the training time in the Unsupervised setting is significantly higher than that of other settings. This is because more training iterations are required for the unsupervised setting, where the amount of pseudo seeds is $7.0-8.2$ times larger than those in the normal setting.
To further study the scalability of \Method{}, we also provide a scalability study based on sampling sub-graphs in Appendix~\ref{app:example_study}.

\begin{table}[t]\small
\caption{The worst and best \HitOne{} when varying $\alpha$}\label{exp:grid_search}
\vspace*{-4mm}
\setlength{\tabcolsep}{1mm}{
\begin{tabular}{lccccccc}
\toprule
\multicolumn{1}{c}{\multirow{2}{*}{Settings}} & \multicolumn{3}{c}{\DICEWS{}}  &  & \multicolumn{3}{c}{\WIKIYAGO{}} \\ \cline{2-4} \cline{6-8} 
\multicolumn{1}{c}{}                          & Worst  & Best & \Method{} &  & Worst  & Best  & \Method{}  \\ \hline
Normal       & 93.7 & \textbf{95.3} & \textbf{95.3} &  & 95.9 & \textbf{98.1} & \textbf{98.1} \\
Less seed    & 90.3 & \textbf{95.3} & \textbf{95.3} &  & 88.4 & \textbf{94.7} & \textbf{94.7} \\
Unsupervised & 94.5 & \textbf{94.6} & \textbf{94.6} &  & 91.0 & \textbf{96.4} & \textbf{96.4}\\
\bottomrule
\end{tabular}
}
\end{table}


\vspace{-2mm}
\section{Conclusions}
\label{sec:conclusions}
In this study, we reinterpret the EA problem for TKGs as a weighted-graph-matching problem and investigate the potential for creating unsupervised methods for EA amongst TKGs by utilizing solitary relation triples with timestamps.
We introduce \Method{}, which unsupervisedly aligns items by fusing relational and temporal data. The proposed \Method{} consists of two steps: (i) encode relational and temporal information independently into embeddings; and (ii) integrate both types of information and transform them into alignment using a novel graph-matching-based decoder. EA on TKGs can be performed using our technique with or without supervision. Furthermore, it does not require additional auxiliary information like entity names or pre-trained word embeddings necessary for earlier unsupervised EA techniques.
Extensive experimental results on several real-world TKG datasets demonstrate that, even in the absence of training data, our technique significantly outperforms the state-of-the-art supervised methods.
In the future, it is of interest to explore dangling settings~\cite{NoMatch21} of EA on temporal datasets. 


\section{Acknowledgements}
This work was supported in part by the National Key Research and Development Program of China under Grant No. 2021YFC3300303, the NSFC under Grants No. (62025206, 61972338, and 62102351), and the Ningbo Science and Technology Special Innovation Projects with Grant Nos. 2022Z095 and 2021Z019.  Lu Chen is the corresponding author of the work.

\balance
\bibliographystyle{ACM-Reference-Format}
\bibliography{REFER}
\pagebreak
\begin{appendices}

\section{Example of Calculation}
We provide a step-by-step example to clarify the calculation process depicted in Figure~\ref{fig:example}. We denote the two TKGs in Figure~\ref{fig:example} as $G_s^\star$ and $G_t^\star$. 
First, we assign unique IDs to entities, relations, and timestamps, as shown in Table~\ref{tab:unique_ids}. Next, we count the outgoing edges by Equation~\ref{eq:time_adj} as well as the incoming edges 
and concatenate matrices.  
For simplicity, we omit the calculation of incoming edges and only demonstrate the process for outgoing edges as follows.

\begin{equation}
    \boldsymbol{A}^t =
\begin{bmatrix}
.40 & .15 & .15 & .15 & .15 \\
.15 & .15 & .40 & .15 & .15 \\
.15 & .40 & .15 & .15 & .15 \\
.20 & .20 & .20 & .20 & .20 \\
.15 & .15 & .15 & .15 & .40 \\
.15 & .15 & .15 & .40 & .15 \\
\end{bmatrix},
\end{equation}

\noindent where $\boldsymbol{A}^t$ is the constructed $\boldsymbol{A}^t$ using $G_s^\star$. We proceed to build the relational adjacency matrix by Equation~\ref{eq:rel_adj}:
\begin{equation}
    \boldsymbol{A}= 
    \begin{bmatrix}
    0 & 0 & 1 & 0 & 0 & 0 \\
    0 & 0 & 1 & 0 & 0 & 0 \\
    0 & 0 & 0 & 1 & 0 & 0 \\
    0 & 0 & 0 & 0 & 0 & 0 \\
    0 & 0 & 0 & 0 & 0 & 1 \\
    0 & 0 & 0 & 0 & 1 & 0 \\
    \end{bmatrix}
\end{equation}

The adjacency matrix $\boldsymbol{A}$ only contains $0$s and $1$s because each relation in Figure~\ref{fig:example} only occurs once. As a result, they do not affect the numerical value of the adjacency matrix.  By setting $L=1$, we obtain the temporal feature $\boldsymbol{H}^t$ which is the concatenation of $\boldsymbol{A}^t$ and $\boldsymbol{AA}^t$ (i.e., $\boldsymbol{H}^t=\left[\boldsymbol{A}^t||\boldsymbol{AA}^t\right]$):
 
\begin{equation}
\boldsymbol{H}^t=
\begin{bmatrix}
.40 & .15 & .15 & .15 & .15 & .15 & .40 & .15 & .15 & .15 \\
.15 & .15 & .40 & .15 & .15 & .15 & .40 & .15 & .15 & .15 \\
.15 & .40 & .15 & .15 & .15 & .20 & .20 & .20 & .20 & .20 \\
.20 & .20 & .20 & .20 & .20 & .00 & .00 & .00 & .00 & .00 \\
.15 & .15 & .15 & .15 & .40 & .15 & .15 & .15 & .40 & .15 \\
.15 & .15 & .15 & .40 & .15 & .15 & .15 & .15 & .15 & .40 
\end{bmatrix}
\end{equation}

\noindent Then, we set $\boldsymbol{A}_s=\boldsymbol{A}_t = \boldsymbol{A}$ and $\boldsymbol{A}^t_s=\boldsymbol{A}^t_t = \boldsymbol{A}^t$ because $G_s^\star$ and $G_t^\star$ in Figure~\ref{fig:example} are identical. This results in normalized WL weights $k^r=1$ and $k^t=1$. Since \RelEncoder{} involves training of GNNs, we randomly initialize an embedding of $|E|\times 10$ for simplicity:

\begin{equation}
    \boldsymbol{H}^r_s = \boldsymbol{H}^r_t=\begin{bmatrix}
.30 & .44 & .33 & .40 & .61 & .35 & .70 & .85 & .32 & .56 \\
.81 & .61 & .97 & .84 & .53 & .00 & .76 & .44 & .19 & .45 \\
.44 & .51 & .70 & .72 & .59 & .21 & .38 & .88 & .90 & .67 \\
.13 & .21 & .77 & .25 & .44 & .71 & .66 & .37 & .35 & .61 \\
.33 & .33 & .70 & .29 & .79 & .51 & .78 & .60 & .24 & .59 \\
.65 & .94 & .77 & 1.00 & .50 & .56 & .50 & .38 & .54 & .34 \\
\end{bmatrix}
\end{equation}

\noindent By setting the weight of \RelEncoder{} to $0$, the alignment matrix $\boldsymbol{P}_T$ is solely determined by the temporal feature:
\begin{equation}
 \boldsymbol{P}_T=\boldsymbol{H}^t(\boldsymbol{H}^t)^T =
 \begin{bmatrix}
.50 & .44 & .39 & .20 & .37 & .37 \\
.44 & .50 & .39 & .20 & .37 & .37 \\
.39 & .39 & .45 & .20 & .39 & .39 \\
.20 & .20 & .20 & .20 & .20 & .20 \\
.37 & .37 & .39 & .20 & .50 & .37 \\
.37 & .37 & .39 & .20 & .37 & .50 \\
\end{bmatrix}
\end{equation}

\noindent With $\boldsymbol{P}_T$, we obtain $\boldsymbol{D}_{G_s^\star G_t^\star}=4.61$ by Equation~\ref{eq:gmd}. Note that, we don't consider Sinkhorn iteration and other processes except dot product for simplicity.

Similarly, by setting the weights of both \TimeEncoder{} and \RelEncoder{} to $1$,  $\boldsymbol{P}$, the alignment matrix obtained by balancing temporal and relational features, is calculated as follows:  

\begin{equation}
     \boldsymbol{P} = \boldsymbol{H}^t(\boldsymbol{H}^t)^T + \boldsymbol{H}^r_s(\boldsymbol{H}^r_t)^T =
\begin{bmatrix}
2.70 & 2.72 & 3.00 & 2.24 & 2.72 & 2.81 \\
2.72 & 3.97 & 3.41 & 2.42 & 2.98 & 3.75 \\
3.00 & 3.41 & 4.03 & 2.59 & 3.02 & 3.67 \\
2.24 & 2.42 & 2.59 & 2.49 & 2.61 & 2.60 \\
2.72 & 2.98 & 3.02 & 2.61 & 3.04 & 2.97 \\
2.81 & 3.75 & 3.67 & 2.60 & 2.97 & 4.26 \\
\end{bmatrix}
\end{equation}

This leads to $\boldsymbol{D}_{G_s^\star G_t^\star}=667.57$, which is significantly larger
than 4.61. This suggests that a lower weight should be assigned to \RelEncoder{}. In this case, as the relational feature was randomly generated, it would be more appropriate to give greater consideration to the temporal feature. Using only \TimeEncoder{} would result in a H@1 of 100\%, whereas using both \TimeEncoder{} and \RelEncoder{} with weights of 1 would decrease H@1 to 66.7\%. This is consistent with the observed distance value.


%

\begin{table}[t]\small
\caption{The ID allocation of Example TKG $G_s^\star$ and  $G_t^\star$}\label{tab:unique_ids}
\vspace*{-4mm}
\setlength{\tabcolsep}{1.mm}{
\begin{tabular}{l|l|c}
\toprule
ID & Name & \multicolumn{1}{c}{Type} \\ \hline
0 & Federal Judge (Argentina) & \multirow{6}{*}{Entities} \\
1 & Nordic Council &  \\
2 & France &  \\
3 & Javier Solana &  \\
4 & Japan &  \\
5 & Head of Government (Ukraine) &  \\ \hline
0 & Reject judicial cooperation & \multirow{5}{*}{Relations} \\
1 & Make an appeal or request &  \\
2 & Sign formal agreement &  \\
3 & Host a visit &  \\
4 & Make a visit &  \\ \hline
0 & 2010-04-24 & \multirow{5}{*}{Timestamps} \\
1 & 2007-09-29 &  \\
2 & 2005-04-16 &  \\
3 & 2007-05-10 &  \\
4 & 2005-07-20 &  \\ \bottomrule
\end{tabular}
}
\end{table}

\section{Dataset Statistics}
\label{app:dataset_stat}
The comprehensive information of all datasets , including \DICEWS{}, \WIKIYAGO{}, and \WIKIYAGOHybrid{} is shown in Table~\ref{tab:datasets}.
\begin{itemize}[topsep=0pt,itemsep=0pt,parsep=0pt,partopsep=0pt,leftmargin=*]
    \item \textbf{\DICEWS{}} is generated from ICEWS05-15~\cite{ICEWS05-15}, which is a subset of facts from the larger ICEWS~\cite{ICEWS} covering the time period from 2005 to 2015. ICEWS05-15~\cite{ICEWS05-15} is widely utilized as a benchmark for TKG comparisons in the community~\cite{TKBC20}. The data generation is followed the method in~\cite{IPTransE17}.  The dataset contains 8,566 entity pairs and all of its facts include temporal information. 
    The overlap ratio of shared quadruples between two TKGs $Q_s$ and $Q_t$ in \DICEWS{} is approximately 50\%, i.e. $2*\frac{|Q_s \cap Q_t|}{|Q_s|+ |Q_t|}\approx 0.5$.
    \item \textbf{\WIKIYAGO{}} is extracted from Wikidata~\cite{Wikidata14} and YAGO~\cite{YAGO}. \WIKIYAGO{} contains 49,172 entity pairs and all of its facts have temporal information. 
    \item \textbf{\WIKIYAGOHybrid{}} is also extracted from Wikidata~\cite{Wikidata14} and YAGO~\cite{YAGO}. However, different from \WIKIYAGO{}, \WIKIYAGOHybrid{ } is a hybrid dataset with 19,462 entity pairs, where 17.5\% of YAGO's facts and 36.6\% of Wikidata's facts do not have temporal information. 
\end{itemize}

\section{Scalability Study on Sampled Datasets}
\label{app:scalability}

To further study the scalability of \Method{}, we follow~\cite{LargeEA22} to sample sub-graphs from \DICEWS{} and \WIKIYAGO{}. 
We vary the number of  entities from 20\% to 100\% and record the running time (seconds) in Table~\ref{exp:scalability_table}, where R, T, and D represent \RelEncoder{}, \TimeEncoder{} and \Decoder{}, respectively. 
 The results indicate that \Method{} is highly scalable. In particular, the running time increases at a quadratic rate as the number of entities increases, which is expected due to the quadratic time complexity of Sinkhorn algorithm. Despite this, even when 100\% of the entities are used, the running time remains relatively short, taking 7 seconds for the \RelEncoder{}, 7.9 seconds for the \TimeEncoder{}, and 13.8 seconds for the \Decoder{} on \DICEWS{} and taking 171.1 seconds for the \RelEncoder{}, 90.2 seconds for the \TimeEncoder{}, and 385.2 seconds for the \Decoder{} on \WIKIYAGO{}. 
 Note that the running time for the different components of \Method{} vary, with the \Decoder{} having the highest running time. This highlights the potential for optimization in the decoding process to further enhance the scalability of the entire \Method{}.

Overall, the scalability study demonstrates that \Method{} is capable of processing large datasets, as it exhibits a relatively low running time even when handling large-scale data. This makes \Method{} well-suited for real-world applications that require efficient and effective data processing.


\section{Open-world Entity Alignment}
Most KGs in the real-world are dynamic, with new entities and timestamps emerging over time~\cite{TREA22}.This makes open-world entity alignment a critical task. However, most existing EA methods adopt a closed-world assumption and are unable to align newly emerging entities\cite{RREA20, DualAMN21, ClusterEA, EASY21, AttrGNN20}. TREA~\cite{TREA22} learns the embeddings of timestamps so that they may satisfy a specific mapping function from actual timestamps, thereby enabling EA on new entities. However, it can only roughly estimate future timestamp embeddings, which may result in errors for subsequent predictions. On the other hand, \Method{} is primarily unsupervised and can be readily adapted to open-world entity alignment  by directly applying the unsupervised component online. As new entities and timestamps become available, \TimeEncoder{} and \Decoder{} can then be re-executed to deliver the most accurate results.
%

To evaluate the open-world effectiveness of EA, we re-arrange the quadruples of \DICEWS{}  according to the open-world setting in TREA~\cite{TREA22}.
The two training sets, $G_s^{\prime}$ and $G_t^{\prime}$, contain quadruples from $G_s$ and $G_t$ in \DICEWS{} covering data prior to Jan. 1, 2014. We train \RelEncoder{} with 1,000 randomly sampled pairs from $G_s^{\prime}$ and $G_t^{\prime}$ following~\cite{TREA22}. The embeddings of unseen entities and relations are  initialized randomly. 


We compare \Method{} with three state-of-the-arts: TREA~\cite{TREA22}, TEA-GNN~\cite{TEAGNN21}, and RREA~\cite{RREA20}. 
The results are shown in Table~\ref{exp:open_world}. \Method{} outperforms the baselines significantly in open-world EA with a $43.2\%$ increase in \HitOne{} for unobserved entity pairs and a $34.9\%$ increase in \HitOne{} for observed entity pairs. This is because both \TimeEncoder{} and \Decoder{} of \Method{} are unsupervised and can be readily deployed online without training.

\section{Case Study}
\label{app:example_study}

We perform a case study of TKG alignment using \Method{}, \Method{}-wt, and \Method{}-wr on \DICEWS{} (denoted as $G_s^\heartsuit$ and $G_t^\heartsuit$), where \Method{}-wt denotes \Method{} without Temporal encoder and \Method{}-wt denotes \Method{} without Relational encoder. Given the ground truth $\{$(Abderrahim, Abderrahim), (Looter, Looter)$\}$, the results delivered by \Method{} are exactly the same as the ground truths. The results generated by \Method{}-wt and \Method{}-wr are detailed as follows.

\newcommand{\Blue}[1]{#1}
\newcommand{\Red}[1]{#1}


\begin{table}[t]\small 
\centering
\caption{Statistics of \DICEWS{}, \WIKIYAGO{}, and \WIKIYAGOHybrid{}}
\vspace{-4mm}
\label{tab:datasets}
\setlength{\tabcolsep}{1.4mm}{
\begin{tabular}{l|c|c|c|c}
\toprule
Dataset & \#Entities  & \#Rels & \#Time & \#Quadruples  \\ \hline
\DICEWS{}       & 9,517-9,537   & 247-246     & 4,017   & 307,552-307,553 \\
\WIKIYAGO{} & 49,629-49,222 & 11-30       & 245    & 221,050-317,814 \\
\WIKIYAGOHybrid{} & 19,493-19,929 & 32-130      & 405    & 83,583-142,568 \\
\bottomrule
\end{tabular}
}
\end{table}

\begin{table}[t]\small
\caption{The result of scalability analysis}
\vspace{-4mm}
\setlength{\tabcolsep}{1mm}{\begin{tabular}{l|lllll|lllll}
\toprule
 & \multicolumn{5}{c|}{\DICEWS{}} & \multicolumn{4}{c}{\WIKIYAGO{}} &  \\ \cline{2-11} 
 & 20\% & 40\% & 60\% & 80\% & 100\% & 20\% & 40\% & 60\% & 80\% & 100\% \\ \hline
 R & 1.1 & 1.5 & 3.1 & 4.6 & 7.0 & 6.0 & 20.9 & 48.5 & 92.1 & 171.1 \\
 T & 0.7 & 1.9 & 3.3 & 5.1 & 7.9 & 2.4 & 10.3 & 26.0 & 52.5 & 90.2 \\
 D & 1.2 & 3.1 & 5.1 & 7.9 & 13.8 & 25.2 & 80.7 & 158.6 & 250.6 & 385.2 \\  \bottomrule
\end{tabular}
}\label{exp:scalability_table}
\end{table}

\begin{table}[t]\small
\caption{The result of open-world entity alignment}\label{exp:open_world}
\vspace*{-4mm}
\setlength{\tabcolsep}{1mm}{
\begin{tabular}{lccclccc}
\toprule
\multicolumn{1}{c}{\multirow{2}{*}{\textbf{Method}}} & \multicolumn{3}{c}{Unobserved Entity Pairs} &  & \multicolumn{3}{c}{Observed Entity Pairs} \\ \cline{2-4} \cline{6-8} 
\multicolumn{1}{c}{} & H@1  & H@10  & MRR   &  & H@1  & H@10 & MRR   \\ \hline
\Method{}                 & \textbf{77.4} & \textbf{89.5} & \textbf{0.829} &  &\textbf{ 94.2 }& \textbf{96.6} & \textbf{0.952} \\
TREA                 & 34.2 & 74.8  & 0.479 &  & 54.9 & 82.5 & 0.643  \\
TEA-GNN              & 15.5 & 59.0  & 0.324 &  & 39.2 & 74.8 & 0.513 \\
RREA                 & 7.5 & 48.3  & 0.253 &  & 36.1 & 58.0 & 0.407
\\\bottomrule
\end{tabular}
}
\vspace*{-1mm}
\end{table}

For entity ``Abderrahim'' in $G_s^\heartsuit$, \Method{} correctly aligns it to  ``Abderrahim'' in $G_t^\heartsuit$, while \Method{}-wr produces an incorrect alignment result ``Chief''. 
The related quadruple of ``Abderrahim'' in $G_s^\heartsuit$ is (Abderrahim, Sign formal agreement,  Foreign, 2007-09-28) and the related quadruple of ``Chief'' in  $G_t^\heartsuit$ is (Chief, Express intent to meet or negotiate, Angola, 2007-09-28). The above case demonstrates that the distinction between the two entities cannot be identified based solely on the timestamp 2007-09-28.


For entity ``Looter'' in $G_s^\heartsuit$, \Method{} correctly align it to  ``Looter'' in $G_t^\heartsuit$, while \Method{}-wt produces an incorrect alignment result ``Rioter''. 
The related quadruple of ``Looter'' in $G_s^\heartsuit$ is (Looter, Use unconventional violence, France, 2005-11-09) and the related quadruple of ``Rioter'' $G_t^\heartsuit$ is 
(Rioter, Use unconventional violence, Police, 2005-03-14). Although $G_s^\heartsuit$ and $G_t^\heartsuit$ have different triples, \RelEncoder{} fails to distinguish them. This is because we only use 10\% of the entity pairs for training the relational model, which can sometimes result in inaccuracies.

In conclusion, \Method{}-wt and \Method{}-wr differ from the ground truths significantly. This implies the necessity of developing and incorporating temporal and relational encoders.

\end{appendices}

\end{document}